\documentclass[twocolumn,amsmath,amssymb,floatfix,superscriptaddress]{revtex4-2}

\ifx\directlua\undefined\ifx\XeTeXcharclass\undefined
  \usepackage[utf8]{inputenc}                           
  \else\RequirePackage[no-math]{fontspec}[2017/03/31]\fi 
  \else\RequirePackage[no-math]{fontspec}[2017/03/31]\fi 
\usepackage{lineno}
\usepackage{siunitx}
\usepackage{tikz}
\usepackage{subfig}
\usepackage{float}

\newcommand{\ifn}{Istituto di Fotonica e Nanotecnologie - Consiglio Nazionale delle Ricerche (IFN-CNR), p.zza Leonardo da Vinci 32, 20133 Milano, Italy.}

\newcommand{\polimi}{Dipartimento di Fisica - Politecnico di Milano, p.zza Leonardo da Vinci 32, 20133 Milano, Italy.}

\begin{document}

\title{High-fidelity and polarization-insensitive universal photonic processors fabricated by femtosecond laser writing}

\author{Ciro Pentangelo}
\affiliation{\polimi}
\affiliation{\ifn}
\author{Niki Di Giano}
\affiliation{\polimi}
\affiliation{\ifn}
\author{Simone Piacentini}
\affiliation{\ifn}
\author{Riccardo Arpe}
\affiliation{\polimi}
\author{Francesco Ceccarelli}
\email[Corresponding author: ]{francesco.ceccarelli@cnr.it}
\affiliation{\ifn}
\author{Andrea Crespi}
\affiliation{\polimi}
\affiliation{\ifn}
\author{Roberto Osellame}
\affiliation{\ifn}

\begin{abstract}
Universal photonic processors (UPPs) are fully programmable photonic integrated circuits that are key components in quantum photonics. With this work, we present a novel platform for the realization of low-loss, low-power and high-fidelity UPPs based on femtosecond laser writing (FLW) and compatible with a large wavelength spectrum. In fact, we demonstrate different UPPs, tailored for operation at \SI{785}{\nano \meter} and \SI{1550}{\nano\metre}, providing similar high-level performances. Moreover, we show that standard calibration techniques applied to FLW-UPPs result in Haar random polarization-independent photonic transformations implemented with average amplitude fidelity as high as 0.9979 at \SI{785}{\nano\metre} (0.9970 at \SI{1550}{\nano\metre}), with the possibility of increasing the fidelity over 0.9990 thanks to novel optimization algorithms. Besides being the first demonstrations of polarization-transparent UPPs, these devices show the highest level of control and reconfigurability ever reported for a FLW circuit. These qualities will be greatly beneficial to applications in quantum information processing.
\end{abstract}

\maketitle



\section{Introduction}
Quantum information processing is a rapidly advancing field that aims at harnessing the unique properties of quantum mechanics, such as superposition and entanglement, to perform computation and communication tasks that are impossible or difficult using classical methods. 
Photonics offers several advantages over other approaches in this framework \cite{flamini2018photonic}. Photons are highly stable and can travel long distances without being absorbed or suffering decoherence even at room temperature. Their flying nature make them also the most natural way to transfer quantum information. Furthermore, interest in this approach has recently increased after the experimental demonstrations of quantum supremacy in photonic systems \cite{zhong2020quantum,madsen2022quantum}.

One promising and scalable approach to implement quantum computing and quantum communication protocols is through the use of photonic integrated circuits (PICs) \cite{pelucchi2022potential}. Integrated photonics allows to miniaturize optical components and integrate them on the same substrate, leading to high scalability and integration density while guaranteeing an intrinsic optical stability even among a large number of components. 
Programmability of the PICs operation is typically achieved by actively controlling the phase shifts
\cite{bogaerts2020programmable}. The simplest and most widely implemented form of phase shifters are thermal phase shifters, which exploit the thermo-optic effect by dissipating electrical power into heat, reversibly modifying the waveguide refractive index.

The simplest fully programmable PIC is the Mach-Zehnder interferometer (MZI), which is a 2-port circuit featuring two balanced directional couplers and two phase shifters. This device can implement any unitary transformation between the input and output modes. The generalization to an $N$-mode circuit can be done by employing a mesh of MZIs in triangular \cite{reck1994experimental} or rectangular \cite{clements2016optimal} configuration, thus obtaining a circuit that is able to perform any unitary transformation in $U(N)$. These universal photonic processors (UPPs) are key components for quantum information processing and have been already demonstrated in various photonic platforms and materials \cite{carolan2015universal,taballione202320mode,de2022high,dong2023programmable,perez2017silicon,harris2017quantum,perez2020multipurpose,tang2021ten,sund2023high,dyakonov2018reconfigurable,kondratyev2023large}. Among them, femtosecond laser writing (FLW) of waveguides in silicate glass \cite{corrielli2021femtosecond} features low insertion losses and low birefringence over a wide wavelength spectrum ranging from the visible to the near-infrared. This fabrication technique is quite versatile: it not only allows for cost-effective and rapid prototyping of PICs, but also enables to ablate the substrate with femtosecond pulses and thus cut out microstructures. The micro-structuring of the substrate allowed by FLW can be used to fabricate thermal isolation structures \cite{ceccarelli2020low} that, in conjunction with thermal phase shifters, significantly reduce their power dissipation and crosstalk of orders of magnitude.

In this work we demonstrate the potential of the FLW platform by fabricating and calibrating two 6-mode UPPs operating at \SI{785}{\nano\metre} and \SI{1550}{\nano\metre}, respectively. These circuits feature insertion losses at \SI{785}{\nano\metre} (\SI{1550}{\nano\metre}) lower than \SI{3}{\decibel} (\SI{2.5}{\decibel}), average $2\pi$ power dissipation per phase shifter as low as \SI{39}{\milli\watt} (\SI{63}{\milli\watt}) and are able to implement unitary transformations with an average amplitude fidelity of 0.9979 (0.9970), which can increase over 0.9990 by exploiting optimization algorithms and which does not depend on the H/V polarization state of the input light. These devices are among the few examples of UPPs currently reported in the literature showing such a high level of control and reconfiguration accuracy over a wide set of implemented transformations and, to the best of our knowledge, the first processors featuring a polarization transparent behaviour.

\begin{figure*}[t]
    \centering
    \input{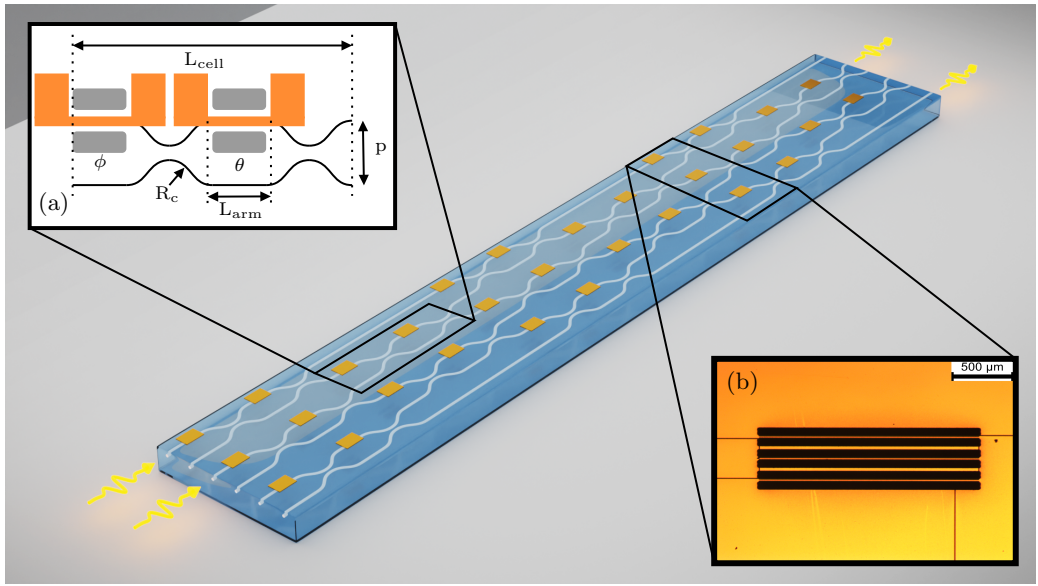}
    \caption{3D rendering of the UPP. Inset (a) shows the schematic layout of an individual MZI of the device. Inset (b) is a microscope picture of UPP A comprising a column of 3 thermal shifters, where it is possible to see the trench structures and the ablations in the metal film.}
    \label{fig:art_attack}
\end{figure*}

\section{Design and fabrication}
Processors at \SI{785}{\nano\metre} and \SI{1550}{\nano\metre} (UPP A and B respectively from now on) share the same waveguide layout based on a rectangular mesh \cite{clements2016optimal} of 15 MZI-based unit cells, entailing a total number of 30 thermal shifters (Figure \ref{fig:art_attack}). The unit cell reported in \cite{ceccarelli2020low} is here employed for UPP A and depicted in Figure \ref{fig:art_attack} (inset a). The pitch between adjacent waveguides is $p =$ \SI{80}{\micro\metre}. Balanced directional couplers are realized by bending the waveguides with a minimum curvature radius of $R_c=$ \SI{30}{\milli\metre}, while MZI arms (and thermal shifters) are $L_{arm}=$ \SI{1.5}{\milli\metre} long. The total length of the cell is $L_{cell}=$ \SI{11.4}{\milli\meter}. This results in a chip dimension of $80 \times$\SI{20}{\milli\meter} including also  the fan-in and fan-out sections at each end of the circuit added for compatibility with standard \SI{127}{\micro\metre} fiber arrays. In order to compensate for the longer operating wavelength and keep the same temperature profile \cite{ceccarelli2020low} for a given phase shift, UPP B instead features longer MZI arms ($L_{arm}=$ \SI{3}{\milli\metre}). Constant-temperature scaling allows us to produce devices sharing the same properties in terms of stability, breakdown power, nonlinearity, etc., paying a small price in terms of unit cell length. However, this penalty is partially compensated by employing more confining waveguides featuring negligible bending losses down to $R_c=$ \SI{15}{\milli\metre}. The reduced radius leads to a total length of the cell $L_{cell}=$ \SI{13.2}{\milli\meter} and, as a result, to a chip dimension of $90 \times$\SI{20}{\milli\meter}. 
\begin{figure*}[t]
    \centering
    \subfloat[\label{fig:int_ramp}]{
        \includegraphics[scale=0.5]{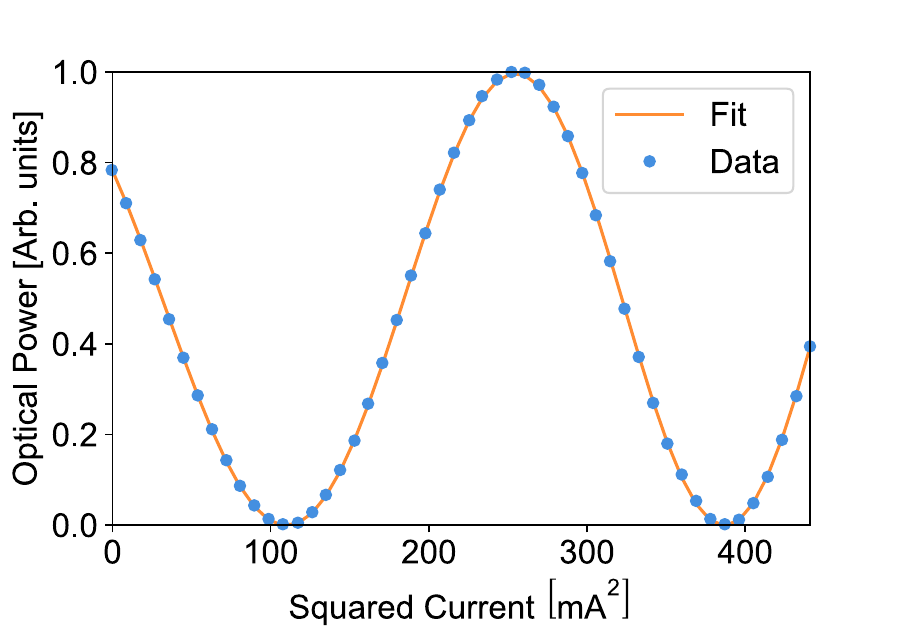}} \quad
    \subfloat[\label{fig:theta_ramp}]{
        \includegraphics[scale=0.5]{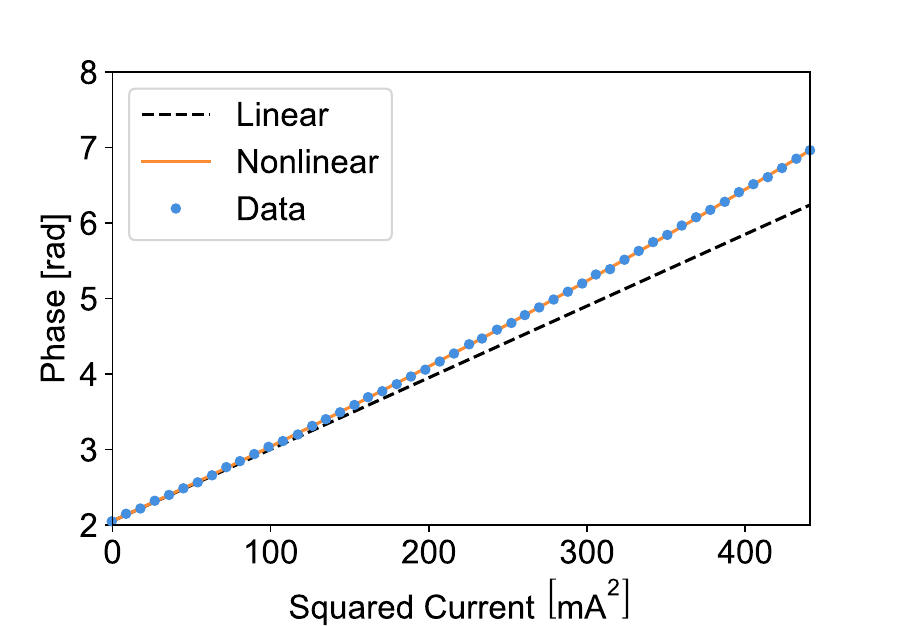}} 
    \caption{Experimental characterization of individual MZIs on UPP A. \protect\subref{fig:int_ramp} Optical power $P_{out}$ measured at the cross output as a function of the squared current $I^2$ when an internal thermal shifter is actuated. Best fit and experimental dataset are both reported, showing the effectiveness of our model, based on Eqs. \ref{eq:ideal_ps} and \ref{eq:nonlinearity}. \protect\subref{fig:theta_ramp} Phase $\theta$ as a function of the squared current $I^2$ obtained from the dataset reported in \protect\subref{fig:int_ramp}. The solid orange line represents the best nonlinear (polynomial) fit. The dashed black line represents the expected trend without the second-order term (i.e. $\beta = 0$ in Eq. \ref{eq:nonlinearity}).}
    \label{fig:interference_fringes_deltaphase}
\end{figure*}

Fabrication of these devices starts from a \SI{1}{\milli\metre} thin Corning Eagle XG alumino-borosilicate glass substrate. Waveguides are inscribed at a depth of \SI{30}{\micro\metre} from the surface, by multi-scan laser irradiation followed by thermal annealing of the substrate \cite{corrielli2018}. Waveguide irradiation parameters are optimized for single-mode operation at the respective wavelengths for the two processors. Thermal isolation trenches are machined by water-assisted laser ablation on each side of the top arm of each MZI both before and after the first directional coupler, where the thermal shifters will be fabricated \cite{ceccarelli2020low}. All trenches are \SI{300}{\micro\metre} deep, \SI{60}{\micro\metre} wide, and either \SI{1.5}{\milli\metre} or \SI{3}{\milli\metre} long respectively for devices A and B. Fabrication of the resistive microheaters of the termal phase shifters is based on the process reported in \cite{ceccarelli2019thermal}. A thin gold layer is deposited on the surface of the device by thermal evaporation and then etched with femtosecond laser pulses so that \SI{10}{\micro\metre} wide microheaters are located on top of the desired MZI arms, while larger contact pads allow for their connection at the sides of the die. A large aspect ratio for the contact pads is required to limit their parasitic series resistance, given that both them and the microheaters are fabricated on the same gold film. Figure \ref{fig:art_attack} (inset b) is a micrograph of UPP A showing a column of three MZI cells, in which it is possible to easily identify trenches, microheaters and contact pads. After packaging the die on an aluminum heat sink, the thermal shifters are connected to printed circuit boards by means of electrically conductive epoxy glue, allowing easy interfacing with the external electronics. Final resistance values for the microheaters are \mbox{111 $\pm$ \SI{6}{\ohm}} (UPP A) and \mbox{215 $\pm$ \SI{15}{\ohm}} (UPP B). Finally, the input and output ports of the circuits are made available for characterization by standard optical fiber arrays pigtailed with UV-curing glue. At the end of this process, total insertion losses of about \SI{3}{\decibel} and \SI{2.5}{\decibel} are measured for UPP A and B, respectively.
\begin{figure*}[t]
    \centering
    \subfloat[\label{fig:A_switch_30_scatter}]{
        \includegraphics[scale=0.5]{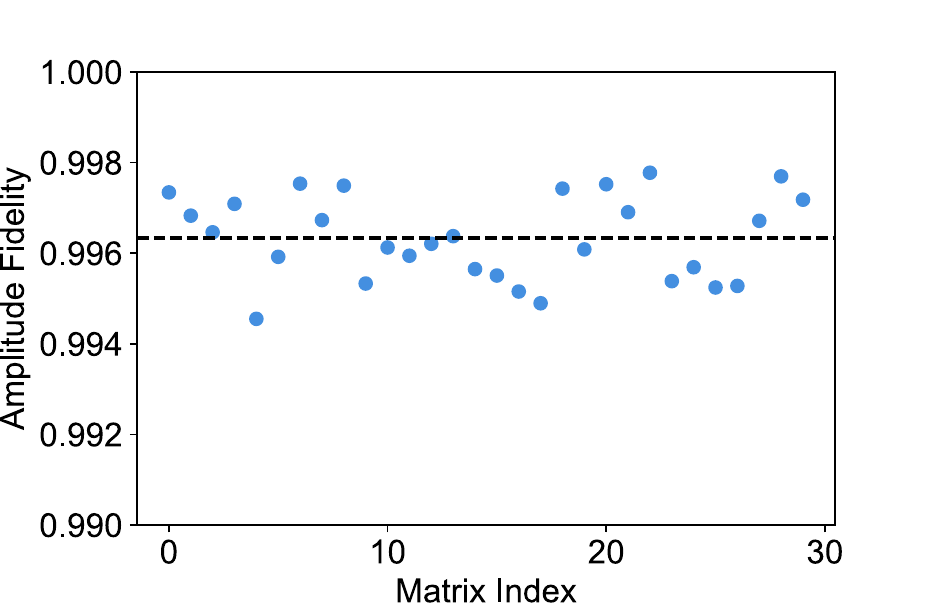}} \quad
    \subfloat[\label{fig:A_comparison_switch_30}]{
        \includegraphics[scale=0.46]{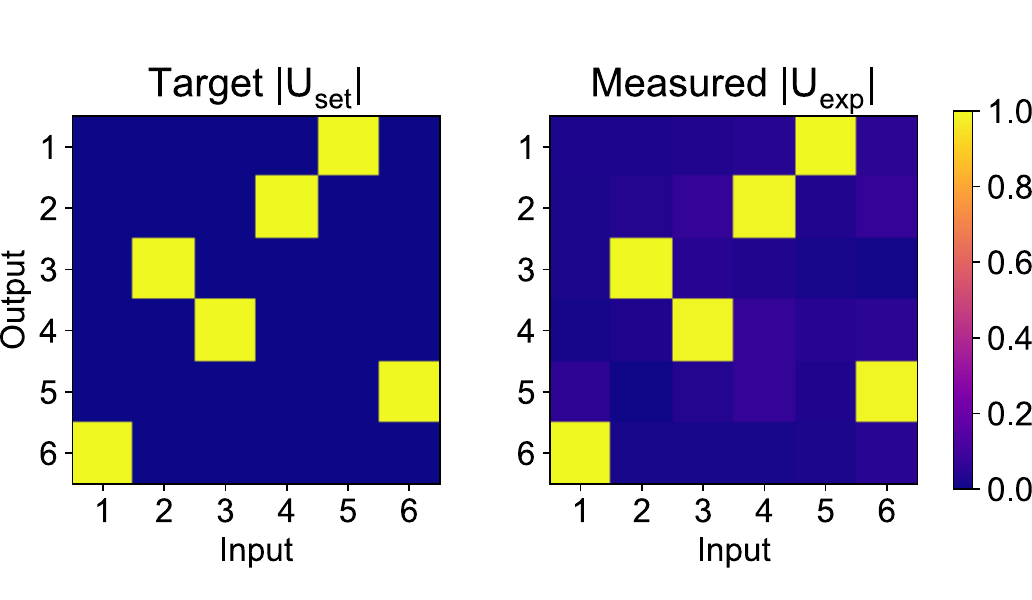}}
\\
    \subfloat[\label{fig:B_switch_30_scatter}]{
        \includegraphics[scale=0.5]{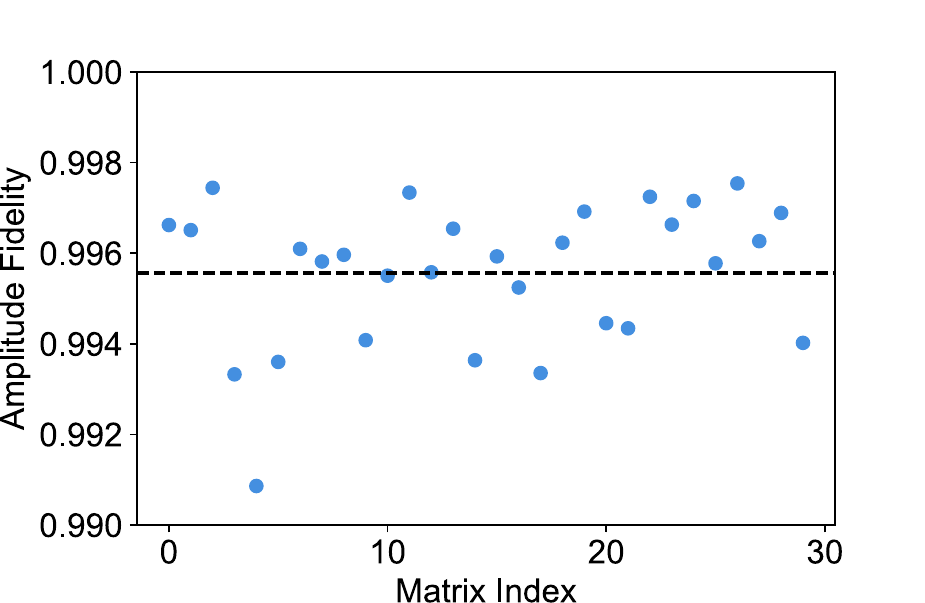}} \quad
    \subfloat[\label{fig:B_comparison_switch_30}]{
        \includegraphics[scale=0.46]{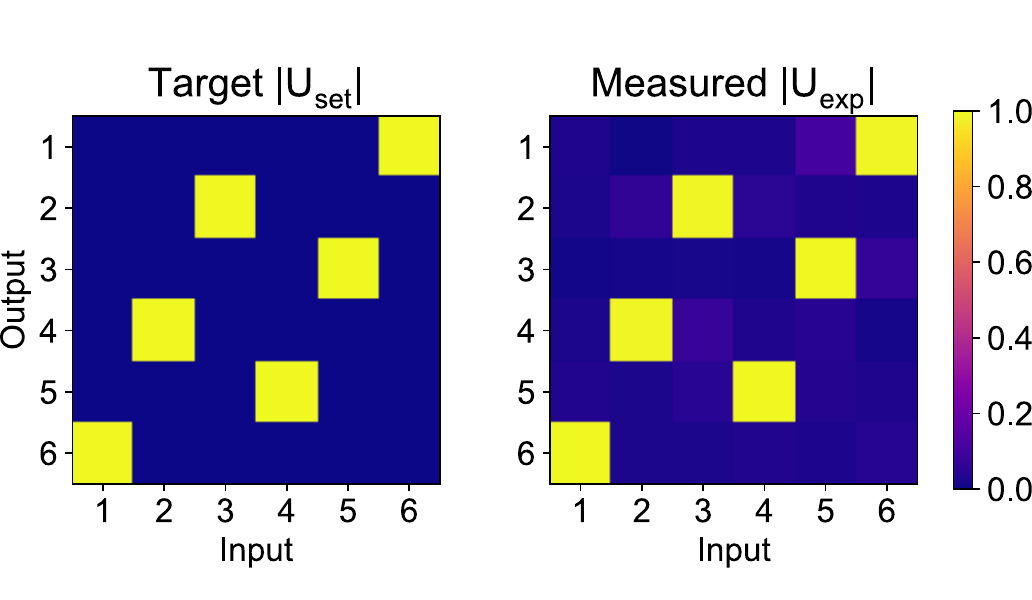}}
    \caption{Amplitude fidelity $\mathcal{F}_{ampl}(U_{set},U_{exp})$ distribution over the 30 randomly chosen switching matrices. \protect\subref{fig:A_switch_30_scatter} Scatter plot of the distribution for UPP A. The average 0.9963 is marked by the dashed line. \protect\subref{fig:A_comparison_switch_30} Example of a switching matrix implementation for UPP A with amplitude fidelity 0.9959. We compare the amplitudes of the target matrix $U_{set}$ versus the amplitudes of the measured matrix $U_{exp}$. \protect\subref{fig:B_switch_30_scatter} Scatter plot of the distribution for UPP B. The average 0.9956 is marked by the dashed line. \protect\subref{fig:B_comparison_switch_30} Example of a switching matrix implementation for UPP B with amplitude fidelity 0.9960. We compare the amplitudes of the target matrix $U_{set}$ versus the amplitudes of the measured matrix $U_{exp}$.}
    \label{fig:switching_results}
\end{figure*}
\begin{figure*}[t]
    \centering
    \subfloat[\label{fig:A_haar_1k_scatter}]{
        \includegraphics[scale=0.5]{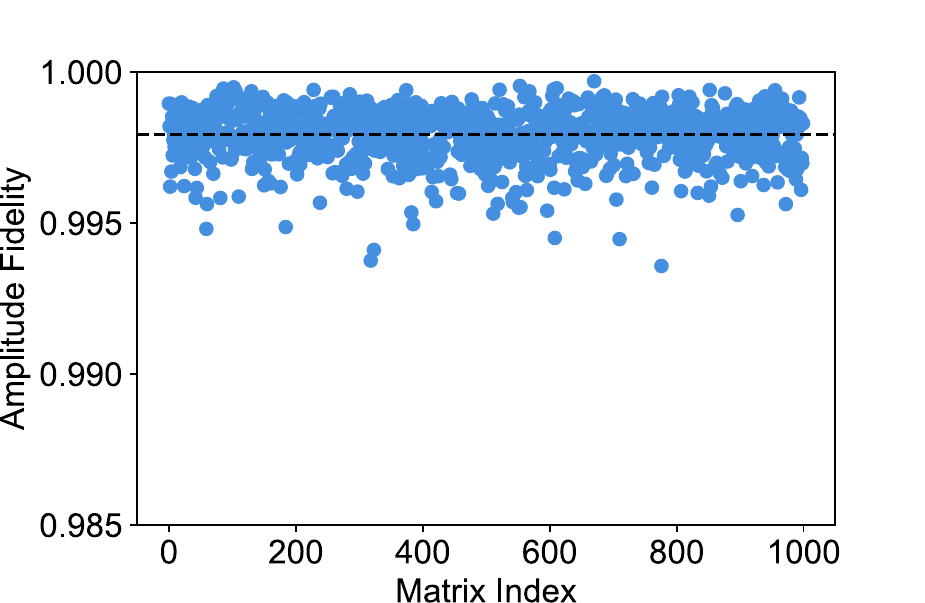}} \quad
    \subfloat[\label{fig:A_comparison_haar_1k}]{
        \includegraphics[scale=0.46]{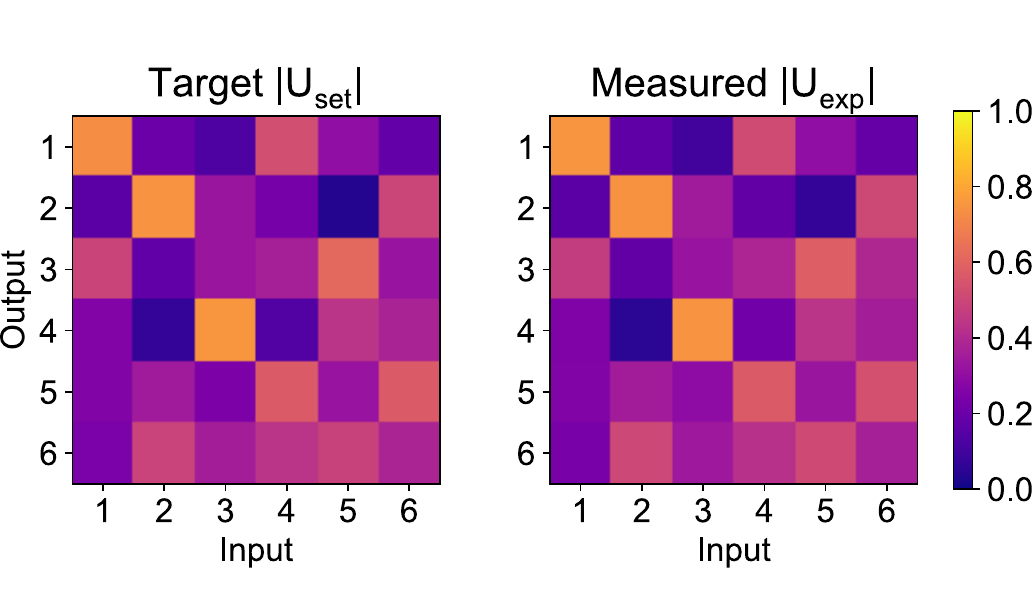}}
    \\
    \subfloat[\label{fig:B_haar_1k_scatter}]{
        \includegraphics[scale=0.5]{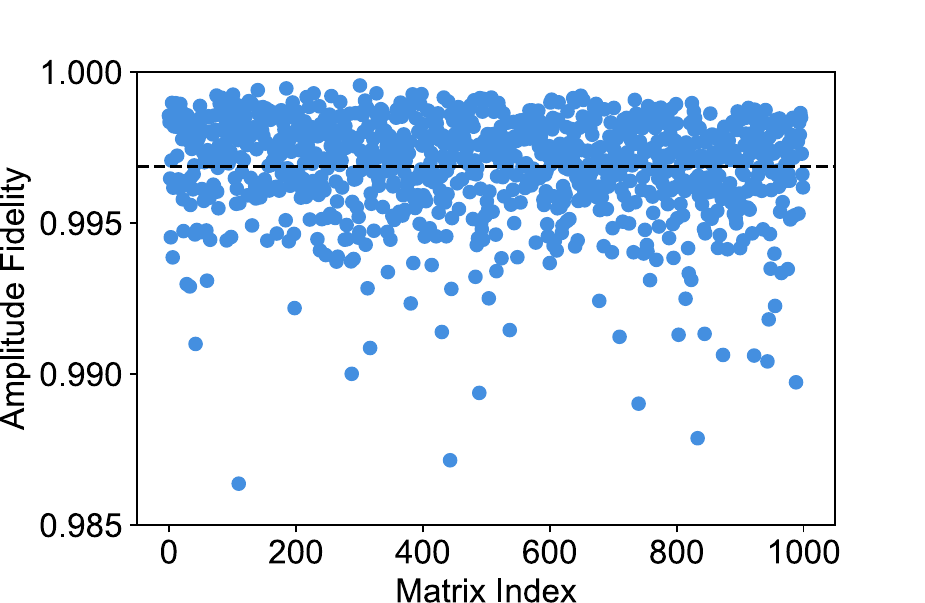}} \quad
    \subfloat[\label{fig:B_comparison_haar_1k}]{
    \includegraphics[scale=0.46]{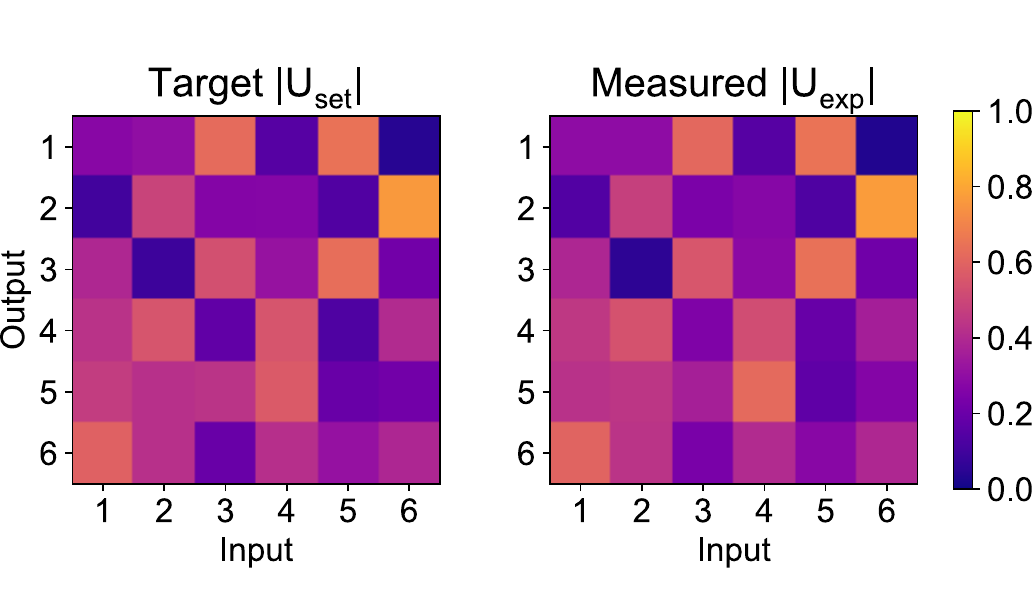}}
    \caption{Amplitude fidelity $\mathcal{F}_{ampl}(U_{set},U_{exp})$ distribution over the 1000 Haar random unitary matrices. \protect\subref{fig:A_haar_1k_scatter} Scatter plot of the distribution for UPP A. The average 0.9979 is marked by the dashed line. \protect\subref{fig:A_comparison_haar_1k} Example of a unitary matrix implementation for UPP A with amplitude fidelity 0.9975. We compare the amplitudes of the target matrix $U_{set}$ versus the amplitudes of the measured matrix $U_{exp}$. \protect\subref{fig:B_haar_1k_scatter} Scatter plot of the distribution for UPP B. The average 0.9970 is marked by the dashed line. \protect\subref{fig:B_comparison_haar_1k} Example of a unitary matrix implementation for UPP B with amplitude fidelity 0.9964. We compare the amplitudes of the target matrix $U_{set}$ versus the amplitudes of the measured matrix $U_{exp}$.}
    \label{fig:haar_results}
\end{figure*}

\section{Modeling and calibration}
The transfer matrix of the MZI unit cell reported in Figure \ref{fig:art_attack} (inset a) can be expressed as:
\begin{equation}
    \label{eq:MZI_T}
    \boldsymbol{U_{MZI}} = e^{i\left(\frac{\theta}{2}+\frac{\pi}{2}\right)} 
    \begin{bmatrix}
    e^{i\phi}\sin\left(\frac{\theta}{2}\right) &  
    \cos\left(\frac{\theta}{2}\right) \\
    e^{i\phi}\cos\left(\frac{\theta}{2}\right) & 
    -\sin\left(\frac{\theta}{2}\right)
    \end{bmatrix},
\end{equation}
where $\phi$ and $\theta$ are the phases induced by the external and internal phase shifters respectively (see Figure \ref{fig:art_attack}, inset a). Assuming to inject light in one input port of this cell, the normalized optical power $P_{\text{out}}$ measured at the cross output port will depend only on the internal phase $\theta$ as:
\begin{equation}\label{eq:ideal_ps}
    P_{out} = \frac{1 + \cos\left(\theta\right)}{2}.
\end{equation}
The phase $\theta$ induced by a thermal shifter can be tuned by controlling either the voltage drop $V$ across the microheater or the current $I$ flowing through it. In our case we have decided for the latter in order to prevent the nonlinear crosstalk due to pure electrical phenomena \cite{ceccarelli2019thermal}. An example of interference measured on an individual MZI is reported in Figure \ref{fig:int_ramp}, where the optical power $P_{out}$ is reported as a function of the squared current $I^2$. Indeed, the phase $\theta$ induced by each shifter can be expressed as follows:
\begin{equation}\label{eq:nonlinearity}
    \theta = \theta_0 + \alpha_I I^2(1 + \beta I^2),
\end{equation}
where the constant phase term $\theta_0$ is an offset present due to fabrication tolerances, $\alpha_I$ is the tuning coefficient of the thermo-optic process and $\beta$ is a correction factor needed to take into account that the microheater resistance depends on the temperature. Such a nonlinear effect is highlighted in Figure \ref{fig:theta_ramp}, where $\theta$ is reported as a function of the squared current $I^2$. In addition, it is also necessary to consider the thermal crosstalk effects. Indeed, the phase induced on the $i$-th MZI in the circuit will be affected by all of the active microheaters and thus:
\begin{equation}\label{eq:crosstalk}
    \theta_i = \theta_{0,i} + \sum_j{\alpha_{ij}I_j^2} (1 + \beta_j I_j^2),
\end{equation}
where the superposition principle is employed in spite of the presence of the correction term thanks to the fact that the latter depends in first approximation only on the $j$-th shifter. In addition, it is worth noting that the constants $\alpha_{ij}$ strongly depend on the distance between the $i$-th MZI and $j$-th shifter. Due to the large bending radii (relative to the inter-waveguide pitch) of these circuits, horizontally neighboring MZIs are millimeters apart while vertically neighboring MZIs are \SI{160}{\micro\metre} apart. This means that we can neglect the coefficients $\alpha_{ij}$ for pairs of MZIs that are not vertically adjacent, leading to a significant simplification of the calibration process and improved control accuracy.

The dataset composed by $\theta_{0,i}$, $\alpha_{i,j}$ and $\beta_j$ represents the calibration dataset for the internal shifters. In order to retrieve it, coherent light is injected in each individual MZI following a node isolation algorithm \cite{Alexiev:21}. Then, the output optical power dependence on the electrical power is fitted from Equations \ref{eq:ideal_ps} and \ref{eq:crosstalk} in order to obtain all the parameters for individual shifters and pairs connected by crosstalk effects. During this process internal shifters that are already calibrated are set to obtain behaviors as straight waveguides ($\theta = \pi$), crossings ($\theta = 0$), or balanced beam splitters ($\theta = \pi/2$). By surrounding a yet uncalibrated MZI with fully reflective or fully transmissive paths, it is possible to isolate it and proceed with a clean characterization of the phase shifter.

For external shifters the procedure follows both the same modeling and measurement strategy. The only difference is the necessity to enclose the phase shifter in larger interferometric rings formed by multiple MZIs \cite{carolan2015universal,harris2017quantum}. All of these measurements have been automated with custom Python scripts to control the instrumentation involved and fit the parameters. More information about the calibration apparatus is reported in the Supplementary Materials (Section S1).

To set a specific unitary transformation $U$ on a UPP one can use the decomposition reported in \cite{clements2016optimal} to obtain the corresponding set of phases $\theta_i$ and $\phi_i$. Then, it is possible to invert Equations \ref{eq:crosstalk} to find the set of currents $I_i$ that implement the desired phases. Since this problem in general does not have a unique solution, we always look for the set of currents $I_i$ that minimizes the total power budget dissipated on chip. With this method, the measured dissipated power was always lower than \SI{1.2}{\watt} in UPP A (\SI{1.9}{\watt} in UPP B). From this calibration procedure it is already possible to estimate the average $2\pi$ power dissipation of each thermal phase shifter, which is \SI{39}{\milli\watt} for UPP A and \SI{63}{\milli\watt} for UPP B.
\begin{figure*}[t]
    \centering
    \subfloat[\label{fig:optimization_scatter}]{
        \includegraphics[scale=0.5]{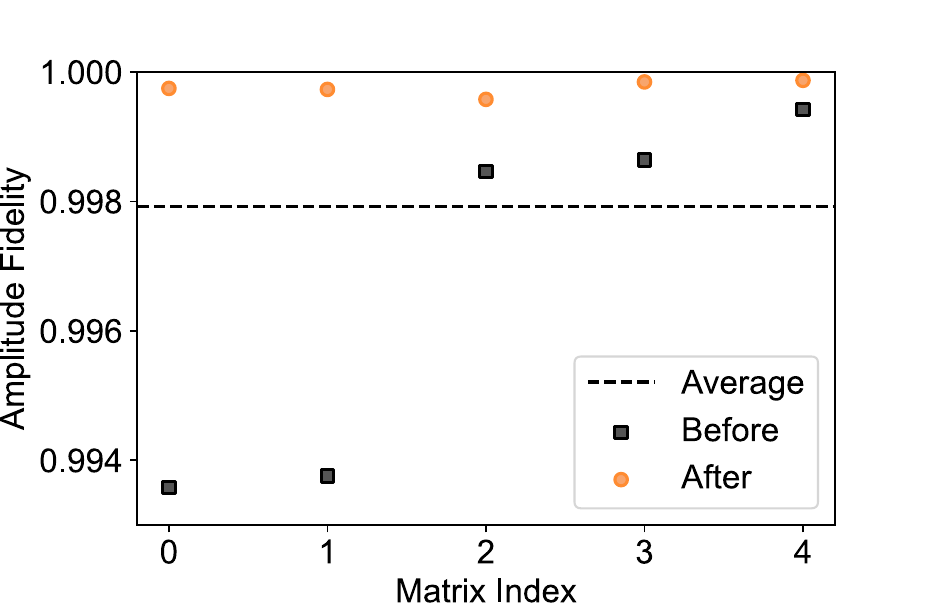}} \quad
    \subfloat[\label{fig:worst_haar_optimized_error}]{
        \includegraphics[scale=0.46]{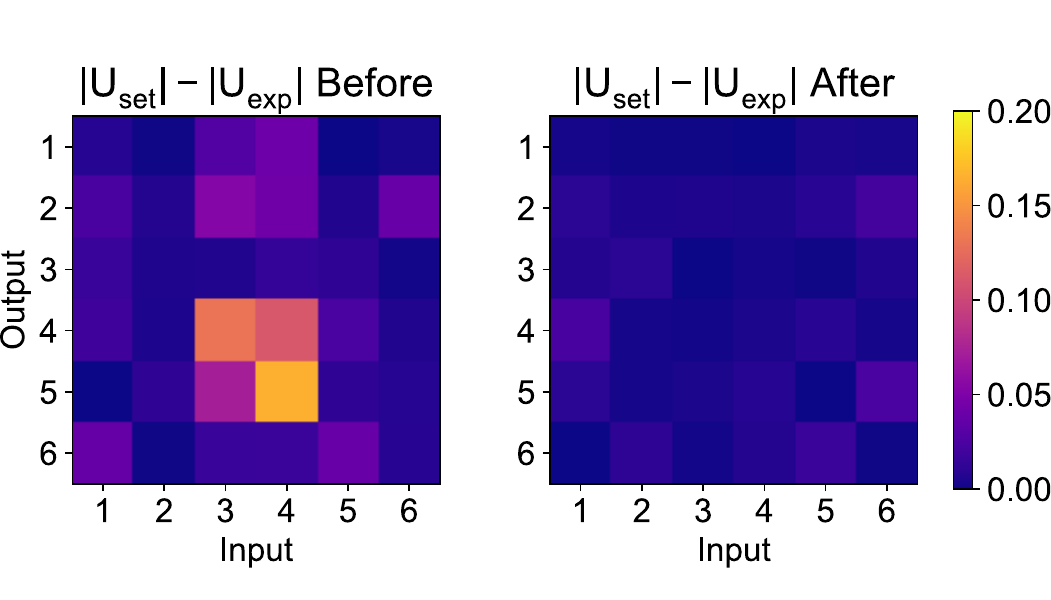}}
    \caption{Amplitude fidelity $\mathcal{F}_{ampl}(U_{set},U_{exp})$ improvement of Haar random matrix implementation through Nelder-Mead algorithm on UPP A. \protect\subref{fig:optimization_scatter} Five unitary matrices (black squares) that were chosen for the optimization and their improved implementation after the optimization process (orange circles). The dashed line is the average fidelity of UPP A over the set of Haar random unitaries as in Figure \ref{fig:A_haar_1k_scatter}. \protect\subref{fig:worst_haar_optimized_error} Difference between the amplitudes of $U_{set}$ and $U_{exp}$ before and after the optimization. This particular matrix was optimized from an amplitude fidelity of 0.9936 to 0.9997.}
    \label{fig:optimization_results}
\end{figure*}

\section{Experimental results}

\subsection{Implementation of unitary transformations} \label{sec:unitaries}
The successful calibration of UPPs A and B was verified with the same experimental setup by implementing two types of unitary transformations: switching transformations, where the device acts as an optical switch linking each input with a given output, and Haar random transformations, corresponding to randomly sampled complex unitary matrices. The former only requires the actuation of internal shifters (specifically to either $\theta = 0$ or $\theta = \pi$) while the latter requires the actuation of both internal and external shifters to arbitrary phase values. Each measurement can be summarized as follows: 
\begin{enumerate}
\item Sample a random switching or Haar random matrix $U_{set} \in U(6)$.
\item Find the set of phases $\theta_i$ and $\phi_i$ corresponding to $U_{set}$ using the decomposition algorithm reported in \cite{clements2016optimal}.
\item Employ the calibration data to extract and implement electrical currents corresponding to desired phases.
\item Measure the input-output intensity distribution and reconstruct the amplitudes of the experimental matrix $U_{exp}$ \cite{hoch2023}.
\item Evaluate the implementation quality by the amplitude fidelity metric (with $N = 6$ being the number of modes):
\begin{equation}
    \mathcal{F}_{ampl}(U_{set},U_{exp}) = \frac{1}{N}tr(|U_{set}^\dagger||U_{exp}|).
    \label{eq:ampl_fid}
\end{equation}
\end{enumerate}

A total of 30 switching unitaries and 1000 Haar random unitaries were implemented on each UPP and the results are summarized in Figure \ref{fig:switching_results} and \ref{fig:haar_results}, respectively.

The amplitude fidelity for the 30 measured switching unitaries is distributed with $\mathcal{F}_{ampl} = \mu \pm \sigma = 0.9963 \pm 0.0009$ for UPP A (Figure \ref{fig:A_switch_30_scatter}) and $0.9956 \pm 0.0016$ for UPP B (Figure \ref{fig:B_switch_30_scatter}). An example of implementation is reported in Figure \ref{fig:A_comparison_switch_30} and \ref{fig:B_comparison_switch_30}, where we compare the amplitudes of the target matrix $U_{set}$ with the reconstructed amplitudes of $U_{exp}$ and we achieve an amplitude fidelity of 0.9959 (UPP A) and 0.9960 (UPP B). These excellent results not only demonstrate the high accuracy that our calibration protocol can reach on the internal phases, but also that the FLW process is able to achieve remarkable accuracy and reproducibility in the implementation of directional couplers with the required splitting ratio.


The amplitude fidelity for the 1000 measured Haar random unitaries is distributed with $\mathcal{F}_{ampl} = \mu \pm \sigma = 0.9979 \pm 0.0009$ for UPP A (Figure \ref{fig:A_haar_1k_scatter}) and $0.9970 \pm 0.0017$ for UPP B (Figure \ref{fig:B_haar_1k_scatter}). An example of implementation is reported in Figure \ref{fig:A_comparison_haar_1k} and \ref{fig:B_comparison_haar_1k}, where we compare the amplitudes of the target matrix $U_{set}$ with the reconstructed amplitudes of $U_{exp}$ and we achieve an amplitude fidelity of 0.9975 (UPP A) and 0.9964 (UPP B). These results demonstrate that the high calibration accuracy reached for the internal shifters was successfully extended also to the external shifters. Universal reconfiguration and high fidelity control is thus demonstrated for both UPPs.
\begin{figure*}[t]
    \centering
    \subfloat[\label{fig:fidelity_t_m_V_H}]{
        \includegraphics[scale=0.5]{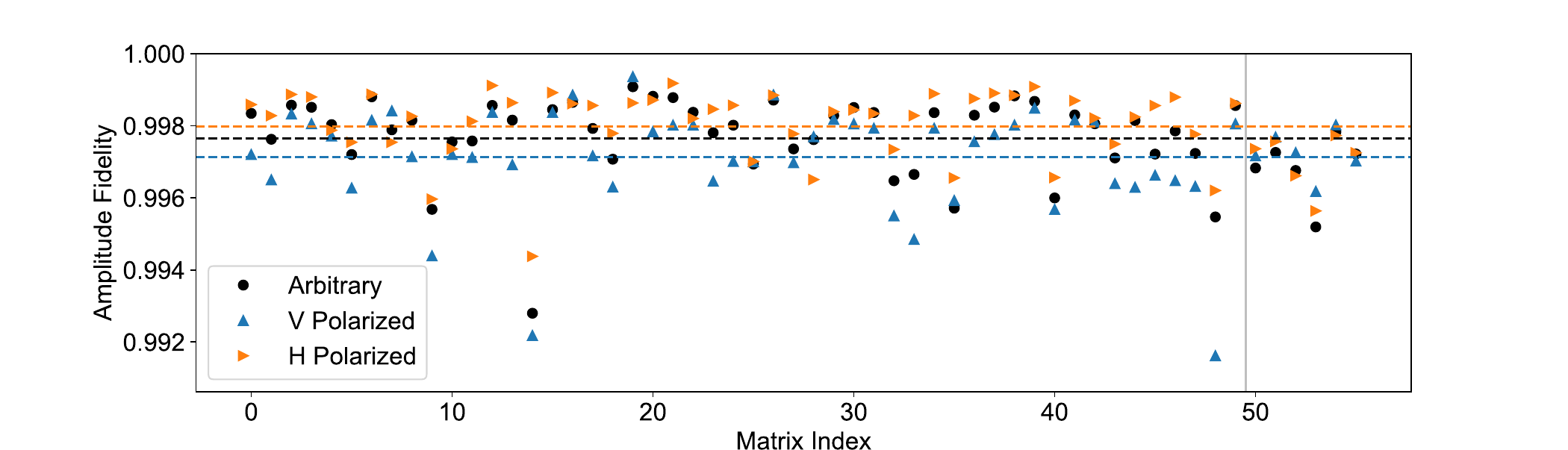}}
    \\
    \subfloat[\label{fig:fidelity_V_H}]{
        \includegraphics[scale=0.5]{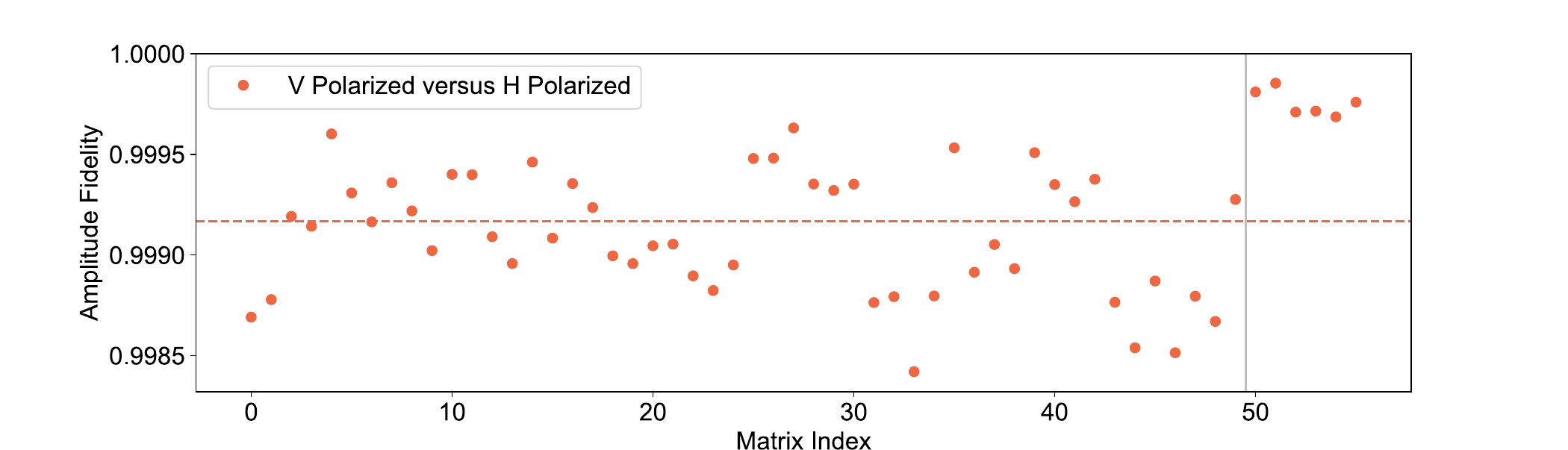}}
    \caption{Amplitude fidelity $\mathcal{F}_{ampl}$ distribution with different polarization states on UPP A. For all these plots, the vertical line separates the set of 50 random Haar matrices from the 6 switching matrices. \protect\subref{fig:fidelity_t_m_V_H} Scatter plot of the amplitude fidelity $\mathcal{F}_{ampl}(U_{set},U_{exp})$ where the experimental matrix $U_{exp}$ was measured for arbitrary as well as V and H polarized light. The averages 0.9978, 0.9971 and 0.9980 are marked by the black, blue and orange dashed lines for the three polarization states, respectively. \protect\subref{fig:fidelity_V_H} Scatter plot of the amplitude fidelity $\mathcal{F}_{ampl}(U_{exp,V},U_{exp,H})$. The average 0.9992 is marked by the dashed line.}
    \label{fig:polarization_fidelities}
\end{figure*}

\subsection{Fidelity improvement via single unitary optimization}
\label{sec:optimization}
After successfully demonstrating the implementation of high-fidelity Haar random unitary transformations on both UPPs, we aim at evaluating whether the employed calibration is indeed reaching the limit in terms of accuracy with which our circuits can implement a given matrix. Namely, we tried to improve the implementation of specific transformations by optimizing the electrical currents used to actuate the microheaters. More in detail, the Nelder-Mead algorithm \cite{NelderMead} was employed using the amplitude "infidelity" $1 - \mathcal{F}_{ampl}$ as a loss function and the phases set on all phase shifts as variables to optimize. The starting point for the optimization is the set of phases obtained for the target unitary with the decomposition algorithm discussed in the previous section \cite{clements2016optimal}. After each step of the optimization algorithm the new phases are converted to electrical currents by using the calibration data, the microheaters are actuated and a new amplitude fidelity is computed from the measurement, which is fed back to the optimizer.

This procedure was applied to 5 of the 1000 Haar random unitaries measured originally on UPP A, selecting some with high, low, or average amplitude fidelity. The results are shown in Figure \ref{fig:optimization_scatter}, where it is clear that even unitaries that were originally measured with high amplitude fidelity can be improved over 0.9995, well above the average for UPP A. A visual comparison between the errors obtained before and after the optimization of a single unitary transformation is shown in Figure \ref{fig:worst_haar_optimized_error}, where the amplitude fidelity increased from 0.9936 up to 0.9997. These results indicate that it is possible to optimize specific unitary transformations in case higher fidelity is required.
In addition, we tried implementing the same unitary repeatedly on the circuit. Over 100 iterations, the average amplitude fidelity between any two measurements of the same unitary transformation is about 0.9998 with this experimental setup. The values reported for the optimized matrices are very close to this limit and, therefore, the current optimization is already the best that we can currently verify. A further optimization will be possible in the future by improving the experimental reproducibility. 

\subsection{Polarization measurement}
In all former measurements the polarization state of light was not controlled. The light polarization at the input of the UPP is determined by the polarization state of light at the output of the laser source and by the action of all the optical elements of the experimental setup. In particular, optical fibers rotate the polarization of light. In performing subsequent measurements, drifts may even have occurred. In the following, we will refer to this as "arbitrary polarization". We now show additional measurements gauging the variation in the performance of UPP A when using an input state of light that has been accurately set as horizontal (H) or vertical (V). The characterization setup for this experiment is the same used before, with the addition of polarizers and waveplates to arbitrary set the polarization state of the coherent light used for the measurements. A complete description of the experimental setup and methods used for this experiment is reported in the Supplementary Materials (Section S1).

As a first step, we sampled a set of 50 Haar random unitary matrices and a randomly chosen set of 6 switching matrices. Then, we implemented each transformation again, measuring the corresponding input-output intensity distribution with controlled H or V polarized light, thus reconstructing the amplitudes of the experimental matrices $U_{exp,H}$ and $U_{exp,V}$. To better discuss how the implementation depends on the H/V polarization, we show these data here in two different ways.

Figure \ref{fig:fidelity_t_m_V_H} shows the amplitude fidelity of the measured matrix calculated against the target matrix $U_{set}$ for all three cases: arbitrary, V and H polarization. The graph shows that the H polarization state performs slightly better on average than the other two, with the V state being the worst overall. Nevertheless, no matrix implementation shows an amplitude fidelity lower than 0.9910 and the average values are 0.9971 for the V polarization and 0.9980 for the H one. This is true not only for the Haar random matrices but also for the switching transformations, providing an additional demonstration of the high polarization transparency of the directional couplers.

Then, Figure \ref{fig:fidelity_V_H} shows how similar the two matrices $U_{exp,V}$ and $U_{exp,H}$ are by reporting the amplitude fidelity calculated between the two. The average value is 0.9992 and no pair below 0.9984 is reported. Again, it is worth noting that these amplitude fidelities were very close to the experimental limit of our characterization setup, which means that even though the polarization definitely plays a role in the correct implementation of the matrices, it does not have as much of an impact for the purposes of implementation as the calibration and operation of the chip.

\section{Discussion}
In this work we evaluated the transformations implemented by our UPPs with classical light and intensity measurements, thus reconstructing only the amplitudes of the complex matrix $U_{exp}$ representing each transformation. Being largely employed in the literature for the benchmarking of UPPs \cite{taballione202320mode,de2022high,dong2023programmable,dyakonov2018reconfigurable,kondratyev2023large}, we selected the amplitude fidelity (see Equation \ref{eq:ampl_fid}) as the figure of merit to measure the accuracy reached by our devices in order to guarantee an easy comparison with the literature. However, this topic deserves a deeper discussion.

\subsection{Amplitude fidelity}
For the sake of clarity, let us start by reporting again the definition of the amplitude fidelity $\mathcal{F}_{ampl}$ for the case of two generic unitary matrices $U = \{u_{ij}\}$ and $V = \{v_{ij}\}$:
\begin{equation}
    \mathcal{F}_{ampl}(U, V) = \frac{1}{N}tr\left(|U^\dagger| |V|\right) = \frac{1}{N} \sum_{i,j} |u_{ij}v_{ij}|.
    \label{eq:ampl_fid_uv}
\end{equation}
Being an average over $N$ scalar products, the amplitude fidelity is a normalized measure of how similar the amplitudes of the two matrices $U$ and $V$ are. Indeed, the amplitude fidelity is equal to 1 if and only if $|U|=|V|$, it is always included in the interval $[0,1]$ and it is directly linked to the amplitude variation matrix $|U| - |V| = \{|u_{ij}|-|v_{ij}|\}$ by the following relation:
\begin{equation}
\begin{split}
    \mathcal{F}_{ampl}(U, V) &= 1 - \frac{1}{2N}\sum_{ij}(|u_{ij}| - |v_{ij}|)^2 = \\
    &= 1 - \frac{1}{2N} \tau^2_{ampl}(U,V),
    \label{eq:ampl_fid_err}
\end{split}
\end{equation}
where we have defined:
\begin{equation}
    \tau^2_{ampl}(U, V) = \sum_{ij}(|u_{ij}| - |v_{ij}|)^2,
    \label{eq:ampl_fid_tsv}
\end{equation}
which is the amplitude total squared variation (TSV) calculated between $U$ and $V$. The analytical proof of Equation \ref{eq:ampl_fid_err} is reported in the Supplementary Materials (Section S2). Although the amplitude fidelity represents an easy way to evaluate the accuracy of a UPP, it is also easy to show that this figure of merit is flawed by a strong bias that reaches its minimum value as $N$ approaches infinity. More specifically, in the Supplementary Materials (Section S2) we prove that:
\begin{equation}
    E[\mathcal{F}_{ampl}(U, V)] \sim \frac{\pi}{4} \text{ as } N\rightarrow \infty,
    \label{eq:ampl_fid_bias}
\end{equation}
where the operator $E[\cdot]$ is the expectation value of the amplitude fidelity calculated over Haar randomly distributed $U,V$. Besides this, the amplitude fidelity is also not suitable to evaluate the performance of a UPP in a multiphoton experiment, since in this case also the angles of the matrix elements play an important role.

\subsection{Fidelity}
Provided that a reconstruction of both amplitudes and angles of each complex matrix element is possible \cite{hoch2023}, the actual device fidelity $\mathcal{F}$ can be evaluated as follows:
\begin{equation}
    \mathcal{F}(U, V) = \frac{1}{N}|tr(U^\dagger V)| = \frac{1}{N} \sum_{i,j} u_{ij}^\dagger v_{ij},
    \label{eq:cplx_fid}
\end{equation}
where we can remove the absolute value since $U$ and $V$ are always known up to a global phase term $e^{i\psi}$ that can be arbitrarily chosen. As an example, a similar figure of merit was employed in \cite{carolan2015universal} thanks to two-photon measurements allowing the reconstruction of the angles. The fidelity represents the normalized Frobenius inner product between $U$ and $V$. Similarly to the amplitude fidelity, it is equal to 1 if and only if $U=V$, it is always included in the interval $[0,1]$ and it is directly linked to the variation matrix $U - V = \{u_{ij}-v_{ij}\}$ by the following relation:
\begin{equation}
\begin{split}
    \mathcal{F}(U, V) &= 1 - \frac{1}{2N}\sum_{ij}(u_{ij} - v_{ij})^2 = \\
    &= 1 - \frac{1}{2N} ||U-V||^2,
    \label{eq:cplx_fid_err}
\end{split}
\end{equation}
where we have defined:
\begin{equation}
    ||U- V||^2 = \sum_{ij}(u_{ij} - v_{ij})^2,
    \label{eq:cplx_fid_nor}
\end{equation}
in which $||U- V||$ is the Frobenius norm calculated on the variation matrix $U - V$. The analytical proof of Equation \ref{eq:cplx_fid_err} is reported in the Supplementary Materials (Section S2), along with the proof that the quantity $||U- V||^2$ is given by two separate contributions:
\begin{equation}
    ||U- V||^2 = \tau^2_{ampl}(U, V) + \tau^2_{angle}(U, V),
    \label{eq:cplx_fid_tsv_ampl_angle}
\end{equation}
where we have defined:
\begin{equation}
    \tau^2_{angle}(U, V) = 4\sum_{ij}|u_{ij}v_{ij}|\sin^2\frac{\angle{u_{ij}}-\angle{v_{ij}}}{2}.
    \label{eq:cplx_fid_tsv_angle}
\end{equation}
The latter is the counterpart of the amplitude TSV and we define it as the angle TSV. Wrapping up the discussion, we can conclude from Equation \ref{eq:cplx_fid_err} and \ref{eq:cplx_fid_tsv_ampl_angle} that: 
\begin{equation}
    \mathcal{F}(U, V) = 1 - \frac{1}{2N}(\tau^2_{ampl} + \tau^2_{angle}). \label{eq:cplx_fid_tsv}
\end{equation}
From Equation \ref{eq:cplx_fid_tsv} it is clear that, since it takes into account also the angle TSV, the fidelity $\mathcal{F}$ is always lower than the amplitude fidelity $\mathcal{F}_{ampl}$ calculated on the same matrix pair $U,V$. Related to this, it is also worth noting that the fidelity $\mathcal{F}$ is a quasi-unbiased figure of merit, in the sense that the bias of the expectation value of the fidelity calculated on Haar randomly distributed unitary matrices $U,V$ vanishes as $N$ approaches infinity. More specifically, in the Supplementary Materials (Section S2) we prove that:
\begin{equation}
    E[\mathcal{F}(U, V)] \sim \frac{\sqrt{\pi}}{2N} \text{ as } N\rightarrow \infty.
    \label{eq:cplx_fid_bias}
\end{equation}

\begin{figure*}[t]
    \centering
    \subfloat[\label{fig:fidelity_simulated_dataset}]{
        \includegraphics[scale=0.5]{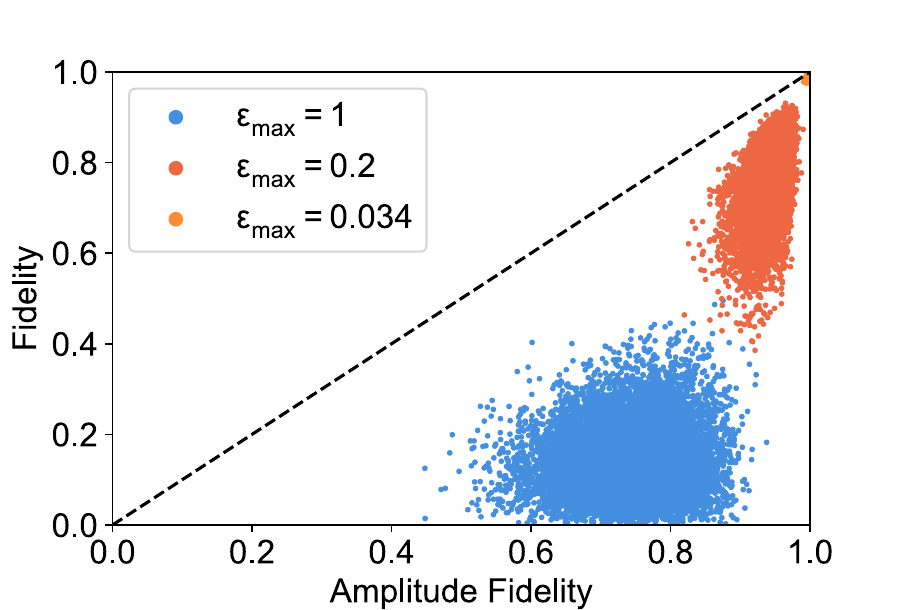}}
    \quad
    \subfloat[\label{fig:optimization_simulated_scatter}]{
        \includegraphics[scale=0.5]{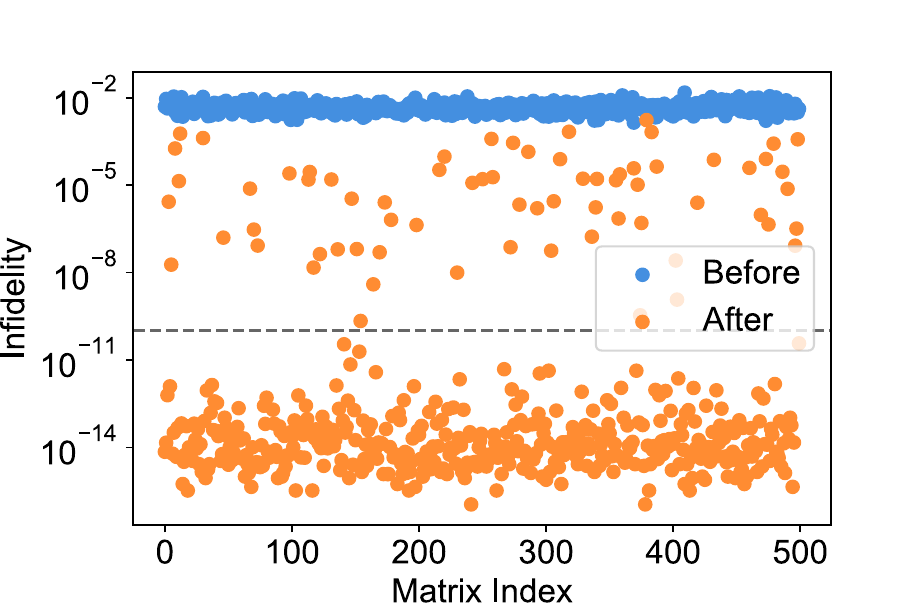}}
    \caption{Simulated scatter plots of fidelity. \protect\subref{fig:fidelity_simulated_dataset} Scatter plot of the simulated fidelities for different values of the random phase noise $\varepsilon_{max}$. The dashed line represents the upper bound of the plot, given by $\mathcal{F} < \mathcal{F}_{ampl}$.
    \protect\subref{fig:optimization_simulated_scatter} Scatter plot of the simulated fidelities after the optimization procedure was performed using the amplitude fidelity as a loss function. The dashed line represents the convergence threshold $1-\mathcal{F} =$ 1E-10.}
    \label{fig:fidelity_simulation}
\end{figure*}

\subsection{Numerical simulation}
Given the doubts raised on the amplitude fidelity $\mathcal{F}_{ampl}$, we decided to implement a Montecarlo simulator to assess the validity of our experimental results.

The simulator goes through the following steps:
\begin{enumerate}
\item Sample a random Haar unitary matrix $U_{set} \in U(6)$.
\item Find the set of phases $\theta_i$ and $\phi_i$ corresponding to $U_{set}$ using the decomposition algorithm reported in \cite{clements2016optimal}.
\item Introduce a random phase noise $\varepsilon$ uniformly distributed in the interval $\varepsilon_{max}[-\pi,\pi]$ on both $\theta_i$ and $\phi_i$.
\item Get $U_{sim}$ by matrix multiplication of each MZI layer.
\item Evaluate the effect of the noise by employing both the amplitude fidelity $\mathcal{F}_{ampl}(U_{set},U_{sim})$ and the actual fidelity $\mathcal{F}(U_{set},U_{sim})$.
\end{enumerate}

Figure \ref{fig:fidelity_simulated_dataset} reports the results of 10000 iterations for three different values of the parameter $\varepsilon_{max}$. For $\varepsilon_{max} = 1$ phases can be considered completely random. Nevertheless, an average amplitude fidelity $\overline{\mathcal{F}_{ampl}} = 0.7411$ is obtained, consistently with the bias that affects this figure of merit. 
On the contrary, the fidelity is a good witness of the high error affecting the phase set, since the simulation produces an average value $\overline{\mathcal{F}} = 0.1475$. For $\varepsilon_{max} = 0.2$ errors are reduced and the amplitude fidelity steeply increases up to $\overline{\mathcal{F}_{ampl}} = 0.9399$. However, the statistical dispersion remains quite large and many unitaries display values lower than 0.9, clearly indicating that something is not working in the processor. In fact, the fidelity remains on average $\overline{\mathcal{F}} = 0.7619$, and several unitaries with very high amplitude fidelity ($\mathcal{F}_{ampl} > 0.95$) have poor fidelity ($\mathcal{F} < 0.6$). Interestingly, the situation looks completely different for $\varepsilon_{max} = 0.034$. This value was chosen to match the amplitude fidelity distribution measured on UPP A both in terms of average and standard deviation, i.e. $\mathcal{F}_{ampl} = 0.9978\pm0.0008$, compared to $\mathcal{F}_{ampl} = 0.9979\pm0.0009$ as reported in Section \ref{sec:unitaries}. In this case, points are all concentrated in a tight spot at the top right corner of the graph in Figure \ref{fig:fidelity_simulated_dataset}; with this phase noise, the fidelity is as high as $\mathcal{F} = 0.9921\pm0.0029$, which is very close to the amplitude fidelity. This suggests that a low statistical dispersion of the amplitude fidelity is a clear witness of low errors also on the angles. These observations strengthen the validity of the experimental characterization performed on our UPPs. Finally, it is worth noting that this simulation allows us also to obtain a rough estimation of the calibration errors, which are evaluated as lower than $\pm$ \SI{0.1}{\radian}.

Secondly, one could also put into question the choice of the amplitude fidelity as a loss function for the optimization process discussed in Section \ref{sec:optimization}. Therefore, we decided to modify the simulator in order to implement the following procedure:
\begin{enumerate}
\item Sample a random Haar unitary matrix $U_{set} \in U(6)$.
\item Find a set of phases $\theta_i$ and $\phi_i$ corresponding to $U_{set}$ using the decomposition algorithm reported in \cite{clements2016optimal}.
\item Introduce a random phase noise $\varepsilon$ suitably distributed to match average and standard deviation of the amplitude fidelity distribution measured for UPP A (Figure \ref{fig:fidelity_simulated_dataset}, orange dots).
\item Apply the minimization algorithm by calculating $U_{sim}$ by matrix multiplication of each MZI layer and by using the amplitude infidelity $1-\mathcal{F}_{ampl}(U_{set},U_{sim})$ as a loss function.
\item Evaluate the final effect of the optimization algorithm by employing the actual infidelity $1-\mathcal{F}(U_{set},U_{sim})$.
\end{enumerate}
Figure \ref{fig:optimization_simulated_scatter} shows the results of 500 iterations of this algorithm, reporting the optimization in terms of infidelity $1-\mathcal{F}$. Despite being based on the amplitude fidelity as the loss function, the algorithm led to a remarkable improvement of the fidelity, with the 86\% of the matrices reaching full convergence (arbitrarily defined as $1-\mathcal{F} <$ 1E-10, dashed line in Figure \ref{fig:optimization_simulated_scatter}). Indeed it is worth noting that no matrix showed a fidelity worse than the initial condition, demonstrating the effectiveness of our optimization protocol based only on intensity measurements.

\section{Conclusion}
In this work, we reported on the design, fabrication and characterization of two 6-mode UPPs fabricated in a FLW integrated photonic platform. Even though larger circuits have been already reported in the literature \cite{kondratyev2023large, hoch2022reconfigurable}, our processors provide the highest level of control and reconfigurability demonstrated to date in a FLW platform, with the additional feature of producing polarization-independent optical transformations. These devices find their natural application in quantum optics and quantum information experiments. The advantages of our technology for this set of applications are manifold. First of all, they are compatible with quantum sources emitting both in the visible range and at telecom wavelength (here demonstrated at \SI{785}{\nano\metre} and \SI{1550}{\nano\metre}) with no penalty in terms of photon losses. Secondly, the high precision reached with our calibration protocol allows for the implementation of arbitrary optical transformations with average fidelity higher than 0.9970, which can be pushed over 0.9990 thanks to an optimization algorithm based only on intensity measurements. Last, the low insertion losses (\mbox{$<$ \SI{3}{\deci\bel}}) make them also compatible with state-of-the-art multiphoton experiments. 

In the future, we believe that the limited power dissipation of our circuits (a few watts), combined with the next generation of thermal phase shifters and programmable MZIs \cite{albiero2022toward}, will enable the scaling towards tens of modes with limited technological effort, thus unlocking a new level of complexity for high-fidelity polarization-insensitive UPPs.

\section*{Acknowledgements}
The authors would like to thank Dr. Simone Atzeni (currently at Paderborn University) for the helpful discussions and the experimental support. Fabrication of the resistive microheaters for the femtosecond laser-written processors was partially performed at PoliFAB, the micro and nano-fabrication facility of Politecnico di Milano \cite{polifab}. The authors would like to thank Emanuele Urbinati (currently at TU Delft) for the help with the fabrication and the PoliFAB staff for the valuable technical support.

\section*{Research funding}
This work is supported by the European Union's Horizon 2020 research and innovation programme under the PHOQUSING project GA no. 899544. R.O. acknowledges funding from the National Centre for HPC, Big Data and Quantum Computing - HPC (CUP B93C22000620006). A.C. acknowledges funding by the PRIN 2017 programme for the Italian Ministry for University and Research, QUSHIP project (Id. 2107SRNBRK).

\section*{Author contributions}
All authors have accepted responsibility for the entire content of this manuscript and approved its submission.

\section*{Conflict of interest}
F.C. and R.O. are co-founders of the company Ephos. The other authors state no conflict of interest.

\section*{Data availability}
The datasets generated and/or analyzed during the current study are available from the corresponding author on reasonable request.

\bibliographystyle{ieeetr}
\bibliography{bib}

\begin{thebibliography}{10}

\bibitem{flamini2018photonic}
F.~Flamini, N.~Spagnolo, and F.~Sciarrino, ``Photonic quantum information
  processing: a review,'' {\em Reports on Progress in Physics}, vol.~82, no.~1,
  p.~016001, 2018.

\bibitem{zhong2020quantum}
H.-S. Zhong, H.~Wang, Y.-H. Deng, M.-C. Chen, L.-C. Peng, Y.-H. Luo, J.~Qin,
  D.~Wu, X.~Ding, Y.~Hu, {\em et~al.}, ``Quantum computational advantage using
  photons,'' {\em Science}, vol.~370, no.~6523, pp.~1460--1463, 2020.

\bibitem{madsen2022quantum}
L.~S. Madsen, F.~Laudenbach, M.~F. Askarani, F.~Rortais, T.~Vincent, J.~F.
  Bulmer, F.~M. Miatto, L.~Neuhaus, L.~G. Helt, M.~J. Collins, {\em et~al.},
  ``Quantum computational advantage with a programmable photonic processor,''
  {\em Nature}, vol.~606, no.~7912, pp.~75--81, 2022.

\bibitem{pelucchi2022potential}
E.~Pelucchi, G.~Fagas, I.~Aharonovich, D.~Englund, E.~Figueroa, Q.~Gong,
  H.~Hannes, J.~Liu, C.-Y. Lu, N.~Matsuda, {\em et~al.}, ``The potential and
  global outlook of integrated photonics for quantum technologies,'' {\em
  Nature Reviews Physics}, vol.~4, no.~3, pp.~194--208, 2022.

\bibitem{bogaerts2020programmable}
W.~Bogaerts, D.~P{\'e}rez, J.~Capmany, D.~A. Miller, J.~Poon, D.~Englund,
  F.~Morichetti, and A.~Melloni, ``Programmable photonic circuits,'' {\em
  Nature}, vol.~586, no.~7828, pp.~207--216, 2020.

\bibitem{reck1994experimental}
M.~Reck, A.~Zeilinger, H.~J. Bernstein, and P.~Bertani, ``Experimental
  realization of any discrete unitary operator,'' {\em Physical Review
  letters}, vol.~73, no.~1, p.~58, 1994.

\bibitem{clements2016optimal}
W.~R. Clements, P.~C. Humphreys, B.~J. Metcalf, W.~S. Kolthammer, and I.~A.
  Walmsley, ``Optimal design for universal multiport interferometers,'' {\em
  Optica}, vol.~3, no.~12, pp.~1460--1465, 2016.

\bibitem{carolan2015universal}
J.~Carolan, C.~Harrold, C.~Sparrow, E.~Mart{\'\i}n-L{\'o}pez, N.~J. Russell,
  J.~W. Silverstone, P.~J. Shadbolt, N.~Matsuda, M.~Oguma, M.~Itoh, {\em
  et~al.}, ``Universal linear optics,'' {\em Science}, vol.~349, no.~6249,
  pp.~711--716, 2015.

\bibitem{taballione202320mode}
C.~Taballione, M.~C. Anguita, M.~de~Goede, P.~Venderbosch, B.~Kassenberg,
  H.~Snijders, N.~Kannan, W.~L. Vleeshouwers, D.~Smith, J.~P. Epping, {\em
  et~al.}, ``20-mode universal quantum photonic processor,'' {\em Quantum},
  vol.~7, p.~1071, 2023.

\bibitem{de2022high}
M.~de~Goede, H.~Snijders, P.~Venderbosch, B.~Kassenberg, N.~Kannan, D.~H.
  Smith, C.~Taballione, J.~P. Epping, H.~v.~d. Vlekkert, and J.~J. Renema,
  ``High fidelity 12-mode quantum photonic processor operating at {InGaAs}
  quantum dot wavelength,'' {\em arXiv preprint arXiv:2204.05768}, 2022.

\bibitem{dong2023programmable}
M.~Dong, M.~Zimmermann, D.~Heim, H.~Choi, G.~Clark, A.~J. Leenheer, K.~J. Palm,
  A.~Witte, D.~Dominguez, G.~Gilbert, {\em et~al.}, ``Programmable photonic
  integrated meshes for modular generation of optical entanglement links,''
  {\em npj Quantum Information}, vol.~9, no.~1, p.~42, 2023.

\bibitem{perez2017silicon}
D.~Perez, I.~Gasulla, F.~J. Fraile, L.~Crudgington, D.~J. Thomson, A.~Z.
  Khokhar, K.~Li, W.~Cao, G.~Z. Mashanovich, and J.~Capmany, ``Silicon
  photonics rectangular universal interferometer,'' {\em Laser \& Photonics
  Reviews}, vol.~11, no.~6, p.~1700219, 2017.

\bibitem{harris2017quantum}
N.~C. Harris, G.~R. Steinbrecher, M.~Prabhu, Y.~Lahini, J.~Mower, D.~Bunandar,
  C.~Chen, F.~N. Wong, T.~Baehr-Jones, M.~Hochberg, {\em et~al.}, ``Quantum
  transport simulations in a programmable nanophotonic processor,'' {\em Nature
  Photonics}, vol.~11, no.~7, pp.~447--452, 2017.

\bibitem{perez2020multipurpose}
D.~P{\'e}rez-L{\'o}pez, A.~L{\'o}pez, P.~DasMahapatra, and J.~Capmany,
  ``Multipurpose self-configuration of programmable photonic circuits,'' {\em
  Nature Communications}, vol.~11, no.~1, p.~6359, 2020.

\bibitem{tang2021ten}
R.~Tang, R.~Tanomura, T.~Tanemura, and Y.~Nakano, ``Ten-port unitary optical
  processor on a silicon photonic chip,'' {\em ACS Photonics}, vol.~8, no.~7,
  pp.~2074--2080, 2021.

\bibitem{sund2023high}
P.~I. Sund, E.~Lomonte, S.~Paesani, Y.~Wang, J.~Carolan, N.~Bart, A.~D. Wieck,
  A.~Ludwig, L.~Midolo, W.~H. Pernice, {\em et~al.}, ``High-speed thin-film
  lithium niobate quantum processor driven by a solid-state quantum emitter,''
  {\em Science Advances}, vol.~9, no.~19, p.~eadg7268, 2023.

\bibitem{dyakonov2018reconfigurable}
I.~Dyakonov, I.~Pogorelov, I.~Bobrov, A.~Kalinkin, S.~Straupe, S.~Kulik,
  P.~Dyakonov, and S.~Evlashin, ``Reconfigurable photonics on a glass chip,''
  {\em Physical Review Applied}, vol.~10, no.~4, p.~044048, 2018.

\bibitem{kondratyev2023large}
I.~V. Kondratyev, V.~V. Ivanova, S.~A. Zhuravitskii, A.~S. Argenchiev, N.~N.
  Skryabin, I.~V. Dyakonov, S.~A. Fldzhyan, M.~Y. Saygin, S.~S. Straupe, A.~A.
  Korneev, {\em et~al.}, ``Large-scale error-tolerant programmable
  interferometer fabricated by femtosecond laser writing,'' {\em arXiv preprint
  arXiv:2308.13452}, 2023.

\bibitem{corrielli2021femtosecond}
G.~Corrielli, A.~Crespi, and R.~Osellame, ``Femtosecond laser micromachining
  for integrated quantum photonics,'' {\em Nanophotonics}, vol.~10, no.~15,
  pp.~3789--3812, 2021.

\bibitem{ceccarelli2020low}
F.~Ceccarelli, S.~Atzeni, C.~Pentangelo, F.~Pellegatta, A.~Crespi, and
  R.~Osellame, ``Low power reconfigurability and reduced crosstalk in
  integrated photonic circuits fabricated by femtosecond laser
  micromachining,'' {\em Laser \& Photonics Reviews}, vol.~14, no.~10,
  p.~2000024, 2020.

\bibitem{corrielli2018}
G.~Corrielli, S.~Atzeni, S.~Piacentini, I.~Pitsios, A.~Crespi, and R.~Osellame,
  ``Symmetric polarization-insensitive directional couplers fabricated by
  femtosecond laser writing,'' {\em Optics Express}, vol.~26, no.~12,
  pp.~15101--15109, 2018.

\bibitem{ceccarelli2019thermal}
F.~Ceccarelli, S.~Atzeni, A.~Prencipe, R.~Farinaro, and R.~Osellame, ``Thermal
  phase shifters for femtosecond laser written photonic integrated circuits,''
  {\em Journal of Lightwave Technology}, vol.~37, no.~17, pp.~4275--4281, 2019.

\bibitem{Alexiev:21}
C.~Alexiev, J.~C.~C. Mak, W.~D. Sacher, and J.~K.~S. Poon, ``Calibrating
  rectangular interferometer meshes with external photodetectors,'' {\em OSA
  Continuum}, vol.~4, no.~11, pp.~2892--2904, 2021.

\bibitem{hoch2023}
F.~Hoch, T.~Giordani, N.~Spagnolo, A.~Crespi, R.~Osellame, and F.~Sciarrino,
  ``{Characterization of multimode linear optical networks},'' {\em Advanced
  Photonics Nexus}, vol.~2, no.~1, p.~016007, 2023.

\bibitem{NelderMead}
J.~A. Nelder and R.~Mead, ``{A simplex method for function minimization},''
  {\em The Computer Journal}, vol.~7, no.~4, pp.~308--313, 1965.

\bibitem{hoch2022reconfigurable}
F.~Hoch, S.~Piacentini, T.~Giordani, Z.-N. Tian, M.~Iuliano, C.~Esposito,
  A.~Camillini, G.~Carvacho, F.~Ceccarelli, N.~Spagnolo, {\em et~al.},
  ``Reconfigurable continuously-coupled {3D} photonic circuit for boson
  sampling experiments,'' {\em npj Quantum Information}, vol.~8, no.~1, p.~55,
  2022.

\bibitem{albiero2022toward}
R.~Albiero, C.~Pentangelo, M.~Gardina, S.~Atzeni, F.~Ceccarelli, and
  R.~Osellame, ``Toward higher integration density in femtosecond-laser-written
  programmable photonic circuits,'' {\em Micromachines}, vol.~13, no.~7,
  p.~1145, 2022.

\bibitem{polifab}
\url{http://www.polifab.polimi.it/}.

\end{thebibliography}

\end{document}